\documentstyle[epsfig,natbib2,natbibmnfix]{mn2e}

\newcommand{\Mpch}{\mbox{ $h^{-1}$ Mpc}}

\newcommand{\be}{\begin{equation}}
\newcommand{\ee}{\end{equation}}

\def\ltsima{$\; \buildrel < \over \sim \;$}
\def\simlt{\lower.5ex\hbox{\ltsima}}
\def\gtsima{$\; \buildrel > \over \sim \;$}
\def\simgt{\lower.5ex\hbox{\gtsima}}

\title{The Peaks Formalism and the Formation of Cold Dark Matter Haloes}

\author[Ludlow \& Porciani] {\parbox{18cm}{
Aaron D. Ludlow$^{\star}$ and 
Cristiano Porciani \\
}\vspace{0.3cm}\\
{Argelander-Institut f\"{u}r Astronomie, Auf dem H\"{u}gel 71,
D-53121 Bonn, Germany}\\
}

\begin{document}

\maketitle 
\begin{abstract}

We use two cosmological simulations of structure formation to study the conditions
under which dark matter haloes emerge from the linear density field. Our
analysis focuses on matching sites of halo collapse to local density maxima,
or ``peaks'', in the initial conditions of the simulations and provides a
crucial test of the central ansatz of the peaks formalism. By identifying
peaks on a variety of smoothed, linearly extrapolated density fields we
demonstrate that as many as $\sim 70$\% of well-resolved dark matter haloes
form preferentially near peaks whose characteristic masses are similar to that
of the halo, with more massive haloes showing a stronger tendency to reside
near peaks initially. We identify a small but significant fraction of haloes
that appear to evolve from peaks of substantially lower mass than that of the
halo itself. We refer to these as ``peakless haloes'' for convenience. By
contrasting directly the properties of these objects 
with the bulk of the proto-halo population we find two clear differences: 1)
their initial shapes are significantly flatter and more elongated 
than the predominantly triaxial majority, and 2) they are, on average, more strongly compressed by tidal forces associated with their surrounding large scale structure. Using the two-point correlation function we show that peakless haloes tend to emerge from highly clustered regions of the initial density field implying that, at fixed mass, the accretion geometry and mass accretion histories of haloes in highly clustered environments differ significantly from those in the field. This may have important implications for understanding the origin of the halo assembly bias, of galaxy properties in dense environments and how environment affects the morphological transformation of galaxies near groups and rich galaxy clusters.

\end{abstract}


\section{Introduction}
\label{intro} 
\renewcommand{\thefootnote}{\fnsymbol{footnote}}
\footnotetext[1]{E-mail: aludlow@astro.uni-bonn.de} 

Within the standard cold-dark-matter cosmology ($\Lambda$CDM)
non-linear clumps of dark matter (referred to as haloes for short) form
hierarchically through the merger and accretion of previous generations of
virialized objects. It is within these haloes that baryons cool and condense
to form galaxies, the visible tracers of the underlying matter distribution
\citep{White1978}. Understanding galaxy formation thus
requires detailed theoretical predictions for the spatial distribution of dark
matter haloes, their abundance as a function of mass, and how these quantities
evolve temporally. 

Attempts to tackle these issues have come from a number of
directions, the most successful of which has been to resort to direct N-body
simulation of non-linear gravitational clustering starting from initial
conditions where linear theory is applicable. 
Although N-body simulations are powerful probes of non-linear structure formation they are notoriously difficult to implement and provide little theoretical insight into the structure formation process. Identifying the precursors of dark matter haloes in the linear density field, for example, remains an unsolved problem. 

As an alternative to N-body simulations, dynamical models for the collapse of 
overdense regions, such as the spherical \citep{Gunn1972} and ellipsoidal
\cite[e.g.,][]{Zeldovich1970,Icke1973,WhiteSilk1979,Hoffman1986,Lemson1993,Bertschinger1994,Bond1996} collapse
models, can be used to physically motivate the conditions under which haloes
emerge from the linear density field. In the so-called peaks formalism, one assumes that gravitationally bound objects
form at local maxima ({\em peaks}) of the linear density field 
(averaged over the appropriate mass scale) which lay above some density threshold.
It is difficult to trace back the exact origin of this model:
some of its features were already discussed in \citet{Doroshkevich1970} while
the whole idea is briefly sketched in \citet{Peebles1980} (\S 26, p. 124).

The formalism gained great popularity after
\citet{Kaiser1984} showed that the statistical properties of overdense
regions can naturally explain the high-clustering amplitude of Abell clusters. 
This was based on the assumption that rich galaxy clusters form from rare high density peaks
in the linear density field, which are much more strongly clustered than the underlying mass distribution
\cite[see also,][]{Politzer1984,Otto1986,Cline1987,Peacock1987,Coles1989,Mo1997,Matsubara1999}

The mathematical development of peaks theory was laid out in considerable 
detail by \cite{Peacock1985} and \cite{Bardeen1986} (hereafter, BBKS). These authors 
studied the properties of local maxima in Gaussian random fields, providing a number of useful statistics, 
such as the number density of peaks of a given height, and their clustering properties.
Extensions to the work of BBKS to non-Gaussian density fields have also been considered 
\citep{Grinstein1986,Matarrese1986,Catelan1988}.

\citet{Blumenthal1984} used the peak model to sketch how galaxy formation
takes place in a universe dominated by cold dark matter.
The number density and internal structure of haloes forming from linear 
density peaks was evaluated by \citet{Hoffman1985}, while
their acquisition of angular momentum during collapse was discussed in a series of papers by
\citet{Hoffman1986}, \citet{Heavens1988}, \citet{Steinmetz1995} and
\citet{Catelan1996}.
Calculations of the halo mass function, accounting for the ``peak-in-peak''
problem, have been presented by \citet{Manrique1995} 
while \citet{Manrique1996,Manrique1998} discussed the implications for halo mass accretion histories.

Recently, new attention has been given to the peaks formalism  
in attempts to model the large-scale clustering properties of haloes in both real \citep{Desjacques2008,Desjacques2010} and in redshift space \citep{DesjacquesSheth2010}.
The peak model has also been used to investigate why the clustering properties
of haloes of a given mass depend on their formation time \citep{Dalal2008a}, the so-called assembly bias.

Over the past couple of decades, the body of assumptions that fashion the peaks formalism have become clear, and their validity is rarely questioned. The peaks formalism appears to work in a statistical sense \cite[e.g.,][]{White1987,Weinberg1990,Park1991a,Park1991b}, yet the question of whether the peak model for galaxy formation actually reproduces the sites of halo collapse is still largely unverified. \citet{Frenk1988} were the first to address this issue. Using $32^3$-particle N-body simulations 
of relatively small volumes (box sizes of $\sim$ 14 \Mpch) these authors concluded that dark matter haloes form preferentially around high peaks identified in the simulations initial density field. Later, \citet{Katz1993a} challenged these conclusions using a simulation ($144^3$ particles) of considerably larger volume ($\sim$ 50 \Mpch{} box), by noting a rather poor correspondence between peak tracers identified in the initial linear density field and the haloes that subsequently form.
A systematic study of proto-haloes (i.e. the regions in the linear density
field that collapse to form haloes at a given time) in a high-resolution
N-body simulation ($256^3$ particles in a $\sim$ (85 \Mpch)$^3$ box) 
was presented by \citet{Porciani2002b}.
These authors concluded that nearly half of galactic sized proto-haloes 
do not contain a linear density peak within their local Lagrangian volume.
If confirmed, these results highlight a general failure of the peak model ansatz since they suggest that a simple one-to-one mapping between haloes and peaks in the smoothed density field does not exist. 

In this paper we revisit these issues using two high-resolution cosmological N-body simulations ($1024^3$ particles). We identify the locations of density peaks in our simulation initial conditions and contrast these directly with the actual sites of halo formation, improving upon the shortcomings of previous work. The remainder of the paper is organized as follows. Following a brief description of our simulations in \S\ref{sec:Nsims}, we describe our main analysis techniques in \S\ref{sec:analysis}; \S\ref{ssec:halomass} provides a definition of halo mass, and an algorithm for locating peaks and matching them to evolved haloes is given in \S\ref{ssec:smoothing} and \S\ref{ssec:peaks_and_haloes}. In \S\ref{sec:results} we describe our main results, which are summarized in \S\ref{sec:conclusions}.

\section{The Simulations}
\label{sec:Nsims}
We study the growth of cosmic structure using two fully cosmological N-body simulations of large scale structure. Both runs start from Gaussian initial conditions and 
assume a flat $\Lambda$CDM cosmological model with numerical parameters consistent with the WMAP 5-year data release \citep[]{Komatsu2009}. These are: $\Omega_M=0.279$, $\Omega_{\Lambda}=1-\Omega_{M}=0.721$, $\sigma_8=0.817$, $n_s=0.96$, and a Hubble constant $H_0\equiv H(z=0)=73$ km s$^{-1}$ Mpc. 

Each run follows the evolution of the dark matter distribution using $1024^3$ particles in a periodic box with fixed side-lengths equal to $l_{box}=150$\Mpch{} and $1200$\Mpch. Initial conditions are generated at $z=70$ (150\Mpch{} box) and $z=50$ (1200\Mpch{} box) using the Zel'dovich approximation. Integrations are performed in comoving coordinates using a lean version of the tree-PM code Gadget-2 \citep[]{Springel2005b}. For this choice of box size, resolution and cosmological parameters the particle masses are 2.433$\times 10^8 h^{-1}$ M$_{\odot}$ and 1.246$\times 10^{11}$ $h^{-1}$ M$_{\odot}$ in the $150$\Mpch{} and $1200$\Mpch{} boxes respectively. By combining the results from the two runs we are able to robustly probe halo abundances for objects with masses $\simgt 7.79\times 10^9 h^{-1} {\rm M}_{\odot}$, as well as gauge the sensitivity of our results to numerical resolution. Further details regarding the simulations, including a thorough discussion of the simulation initial conditions, can be found in \citet{Pillepich2010}.

\section{Analysis}
\label{sec:analysis}

In this section we provide a brief discussion of our main analysis techniques. This includes a definition of halo mass; a method for generating smoothed density fields, $\delta_s$; a choice of filter function, and an algorithm for locating peaks in $\delta_s$.

\subsection{Definition of halo mass}
\label{ssec:halomass}

Dark matter haloes are identified at $z=0$ using a standard friends-of-friends (FOF) algorithm \citep[]{Davis1985}. Our halo finder has one free parameter, $\ell$, which implicitly defines the linking length in terms of the mean particle density, $n$: $\ell n^{-1/3}$. For well resolved objects, the FOF method selects regions enclosed by an isodensity contour $\propto 1/\ell^3$. In what follows we shall adopt $\ell = 0.2$, the conventional choice for $\Omega=1$ cosmologies. For each halo identified at $z=0$ we compute the FOF mass, defined as the total mass associated with the FOF group, and retain the identity of the most bound particle. When necessary, and at $z=0$, we identify the position of the most bound particle with the halo center.

One advantage of the FOF method over other halo identification techniques is that it imposes no shape restriction on the boundary of the system; particles that lie within one linking length from any other particle are grouped together into a single system, regardless of its final dynamical state. A drawback, however, is the tendency of the FOF method to occasionally link separate objects in transient dynamical states, such as fly-by's or ongoing mergers \cite[e.g.][]{Gelb1994,Governato1997}, which can lead to the appearance of massive haloes through the temporary association of two or more low-mass objects. 

We assess the importance of this effect by computing, for each FOF halo, the distance between its true center, $\mathbf{r}_{\rm mb}$ (defined by the location of its most bound particle) and the geometric center of all FOF particles: $d_{\rm off}=|{\mathbf{r}}_{\rm mb}-{\mathbf{r}}_{\rm CM}|/r_{200}$. As discussed by \citet{Bett2007} and \citet{DOnghia2007}, the ``center-of-mass offset'' is a simple test of the prevalence of unrelaxed substructure in the halo, and can be used to cull the halo catalogs of objects far from equilibrium. We have verified that the results presented in \S\ref{ssec:masscomp} are not particularly sensitive to this selection criterion by repeating the analysis for haloes with $d_{\rm off}<0.1$, and $d_{\rm off}<0.3$. Because of this we shall consider all haloes, regardless of the details of their equilibrium state, in the analysis that follows.

\begin{figure}
\begin{center}
\resizebox{8.5cm}{!}{\includegraphics{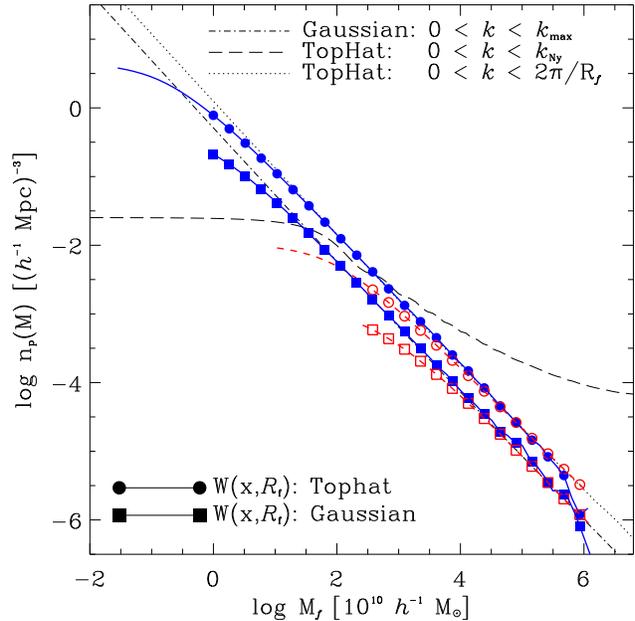}}
\end{center}
\caption{Comoving number density of peaks identified in our simulation initial conditions plotted as a function of filter mass. Peaks are identified on a sequence of linearly extrapolated density fields smoothed with a spherical filter containing mass M$_f$ (see text for details). Curves extend from M$_f=1.16\times 10^{16} h^{-1} {\rm M}_{\odot}$, the maximum smoothing scale considered, down to M$_f={\rm M}_{\rm part}$, corresponding to the unsmoothed density field. Points are over-plotted to highlight the mass range M$_f\geq 32\times {\rm M}_{\rm part}$, the minimum halo mass identified by our group finder. Squares and circles differentiate results obtained with a Gaussian and tophat filter, respectively. The tophat filter contains a characteristic mass M$_f=(4\pi/3) {\rm R}_f^3 \overline{\rho}$; the Gaussian filter has M$_f=(2\pi)^{3/2}{\rm R}_f^3 \overline{\rho}$. Red curves (with open symbols) correspond to our $1200$\Mpch{} box; results from our 150\Mpch{} box are shown in blue. The theoretical curves are the peak number densities expected from peaks theory (eq.~\ref{eq:peaknumdens}) for several different filter choices (see text for details).}
\label{Fig:PeakNumDens}
\end{figure}

\subsection{Identification of linear density ``peaks''}
\label{ssec:smoothing}

For each of our simulations we compute the linear overdensity field $\delta({\mathbf{x}})$ using cloud-in-cell interpolation; densities are computed, in each case, on a $1024^3$ grid. The linear overdensity field can then be smoothed on some spatial scale ${\rm R}_f$ by means of convolution with an appropriate window function, $W(\mathbf{x},{\rm R}_f)$. In what follows we mainly restrict our discussion to results obtained with a real-space tophat filter, but have verified that our conclusions are not particularly sensitive to this choice by repeating parts of the analysis with a Gaussian kernel. 

For small perturbations in the smoothed field, $\delta_s\equiv \delta({\mathbf{x}},{\rm R}_f)$, the tophat filter contains a characteristic mass M$_f\approx (4\pi/3) {\rm R}_f^3 \overline{\rho}$ within ${\rm R}_f$\footnote{Strictly speaking, the filter radius and mass are related by ${\rm M}_f\propto (1+\delta){\rm R}_f^3$, so that for linear perturbations in $\delta$ (i.e., $\delta<<1$) we have ${\rm M}_f\propto {\rm R}_f^3$}. For each of our runs we compute $\delta_s$ in equally spaced steps in $\log {\rm M}_f$ of fixed logarithmic width $\Delta\log {\rm M}_f=0.129$. We choose a minimum filtering mass equal to the particle mass and perform the convolution up to a maximum mass of ${\rm M}_f=1.16\times 10^{16} \ h^{-1}$ M$_{\odot}$. This results in 60 steps in ${\rm M}_f$ for our 150 $h^{-1}$ Mpc box; for our $1200$ $h^{-1}$ Mpc box, which has a higher particle mass, there are a total of 40 steps in ${\rm M}_f$. 

Given the smoothed density fields, $\delta_s$, we next identify local maxima, or ``peaks'', by locating grid cells that are denser than all 26 neighboring points. Peaks are identified on all smoothed fields and their evolution followed by tagging the particle at ${\mathbf{x}}_p$ nearest to each peak grid point. For each peak particle we calculate the associated overdensity, $\delta_p=\delta_s({\mathbf{x}}_p,{\rm R}_f)$, the peak height, $\nu_p=\delta_p/\sigma(\rm{R}_f)$, and the properties of the halo (if any) to which the particle belongs. In the following we consider {\it all} peaks in $\delta_s$ and avoid imposing an arbitrary minimum height for their selection.

With this information we plot in Figure~\ref{Fig:PeakNumDens} the comoving number density of peaks as a function of smoothing scale, $n_p({\rm M}_f)$, for two different choices of filter function. Open and closed symbols distinguish results obtained from our 1200\Mpch{} and 150\Mpch{} boxes, respectively. For the tophat filter points span the mass range $32\times M_{\rm part}< M_f< 10^{16} h^{-1} M_{\odot}$; points are connected by lines which extend down to the particle mass limit corresponding to the unsmoothed density field. Squares highlight $n_p({\rm M}_f)$ obtained for a Gaussian filter and are plotted for filter masses larger than the minimum halo mass identified in each box.

The quantity $n_p({\rm M}_f)$ can also be calculated analytically from the statistical properties of local maxima for a Gaussian field (BBKS). The comoving number density of peaks of arbitrary height is given by 
\begin{equation}
n_p({\rm M}_f)=\frac{29-6\sqrt{6}}{5^{3/2}2(2\pi)^2 R_{\star}^3},
\label{eq:peaknumdens}
\end{equation}
where $R_{\star}\equiv \sqrt{3}(\sigma_1/\sigma_2)$ is the characteristic comoving length-scale of the peaks, and the $\sigma_i$s are spectral moments of the (smoothed) linear power spectrum $P(k)$:
\begin{equation}
\sigma_i^2\equiv \frac{1}{2\pi^2}\int_{0}^{k_{\rm max}} \hat{W}({k},{\rm M}_f)^2 P(k) k^{2(i+1)} dk.
\label{eq:spectralmoments}
\end{equation}

The three curves in Figure~\ref{Fig:PeakNumDens} show the predicted comoving number density of peaks for various choices of filter function and upper limit of integration in eq.~(\ref{eq:spectralmoments}). The dot-dashed curve assumes a Gaussian window function and integrates the full linear power spectrum used to generate the initial conditions of the simulations. The long-dashed curve shows the effect of smoothing with a tophat filter and truncating the integration at the Nyquist frequency of the 1200 $h^{-1}$ Mpc box 
\footnote{The truncation is necessary to avoid the divergence of $\sigma_2$ 
as $k_{\rm max}\to \infty$ for a tophat filter, since for this choice $\hat{W}({k},{\rm M}_f)^2\propto k^{-6}$.}.
In this case, the predicted number density of peaks at large filter masses exceeds that measured directly from the smoothed density fields by a significant amount. For example, at a filtering scale of $\approx 10^{15}$ $h^{-1} {\rm M}_{\odot}$ one expects roughly an order of magnitude more peaks than are actually found in the linear density field.

The origin of this discrepancy is made clear by considering the effect of truncating the integral in eq.~(\ref{eq:spectralmoments}) at $k_{\rm max}=2\pi/{\rm R}_f$ (shown as a dotted-line in Figure~\ref{Fig:PeakNumDens}). This truncation is chosen explicitly in order to ignore the contribution to $n_p({\rm M}_f)$ from power on spatial scales below the characteristic size of the filter. Although this effectively changes the tophat filter function by truncating it in $k-$space, we have verified that imposing the same restriction on the filter prior to smoothing the initial density field of the simulations yields peak number densities consistent with what is expected from eq. ~(\ref{eq:peaknumdens}). In the remainder of the paper we limit our discussion to results obtained with the full tophat filter.   

\begin{figure}
\begin{center}
\resizebox{8.5cm}{!}{\includegraphics{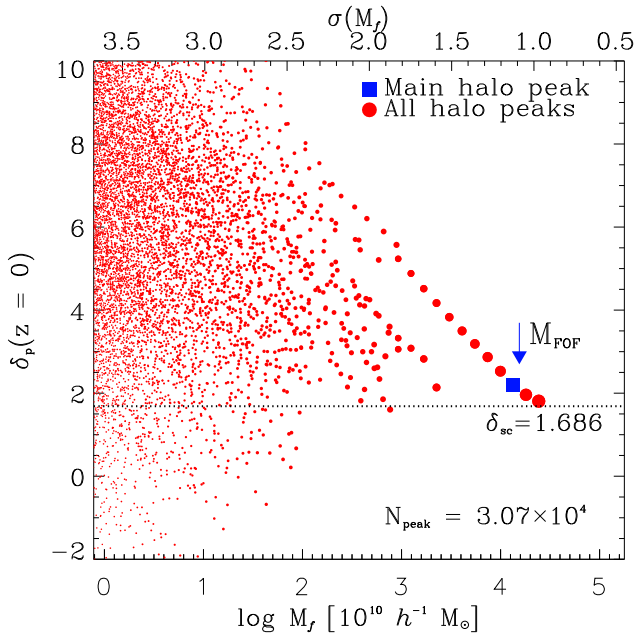}}
\end{center}
\caption{Smoothed linear overdensity, $\delta_p$, extrapolated to $z=0$ for peak particles belonging to a single halo plotted as a function of smoothing scale. Red points denote $\delta_p({\rm M}_f)$ for all peak particles associated with the object, defined as those that end up in the halo by $z=0$. The solid (blue) square shows the peak identified on the smoothing scale nearest to the actual mass of the halo, with the latter shown as a downward pointing arrow. For cosmetic purposes we have randomly perturbed the peak masses by one smoothing bin width for all peaks identified on scales ${\rm M}_f< 10^{13} h^{-1} {\rm M}_{\odot}$.}
\label{Fig:PeakHistory}
\end{figure}

\subsection{Associating density peaks with collapsed haloes}
\label{ssec:peaks_and_haloes}

Having identified peaks on the sequence of smoothed density fields, $\delta_s$, we next associate peak particles with haloes in our friends-of-friends catalogs. This is achieved by scanning the subset of particles belonging to each halo for matches in the list of peaks. By performing the matching at each smoothing scale we thereby build lists of peaks associated with each object. The outcome of applying this procedure to the most massive halo in our 150 $h^{-1}$ Mpc box is shown in Figure~\ref{Fig:PeakHistory}. Here red points show the linearly extrapolated peak overdensity, $\delta_p$, as a function of smoothing scale for all peak particles associated with the halo. 

Clearly well resolved objects will contain an abundance of peak particles with a variety of characteristic masses. In the context of peaks theory, the majority of these peaks correspond to the progenitors from which the halo formed. Although these peak tracers may be of interest for studying the distribution and dynamics of an early generation of accreted (and possibly disrupted) sub-systems \citep[]{Diemand2005d}, the vast majority cannot be responsible for the collapse of the halo as a whole. If one is to trust the primary tenet of the peaks formalism then a single peak, identified on the characteristic smoothing scale ${\rm M}_f \approx {\rm M_{FOF}}$, should be singled out. For the example in Figure~\ref{Fig:PeakHistory} this peak is shown as a solid blue square, and the actual halo mass as a downward pointing arrow.

The results presented in the following Sections are based on the association of a single and unique peak with each FOF halo, which we refer to as the ``main'' halo peak for convenience. For each object we define the main peak as the one identified on the smoothing scale for which the filter mass, ${\rm M}_f$, is nearest to the true halo mass, that also ends up in the halo at $z=0$. This method of associating haloes with peaks in the smoothed linear density field provides a robust lower limit to the fraction of haloes of mass M that are associated with peaks in the density field when smoothed on same characteristic scale, and thus provides a strong test of the primary assumption of the peaks formalism.

One shortcoming of this method was highlighted by \citet{Katz1993a}, who found that the complex merger histories of some haloes may lead to the ejection of peak particles during the formation process. Since our runs lack the temporal resolution to properly track these events, we asses this effect in the following way. If at $z=0$ a halo contains no peak particle from the appropriately filtered linear density field (the one with ${\rm M}_f\approx {\rm M_{FOF}}$), we trace the halo particles back to the initial redshift and search for one such peak within one smoothing radius $R_f=(3{\rm M_{FOF}}/4\pi\overline{\rho})^{1/3}$ from the proto-halo's center of mass. If a peak particle is found to lie within ${\rm R}_f$ at the initial redshift then we attribute the formation of that halo to the peak, and include the particle in the peak catalogs. This typically increases the fraction of collapsed haloes that are properly identified with peaks in the smoothed density field by $5-10$\%, with a slight tendency to be higher for lower mass objects. 

\section{Results}
\label{sec:results}

\subsection{Peak versus halo masses}
\label{ssec:masscomp}

\begin{figure}
\begin{center}
\resizebox{8.5cm}{!}{\includegraphics{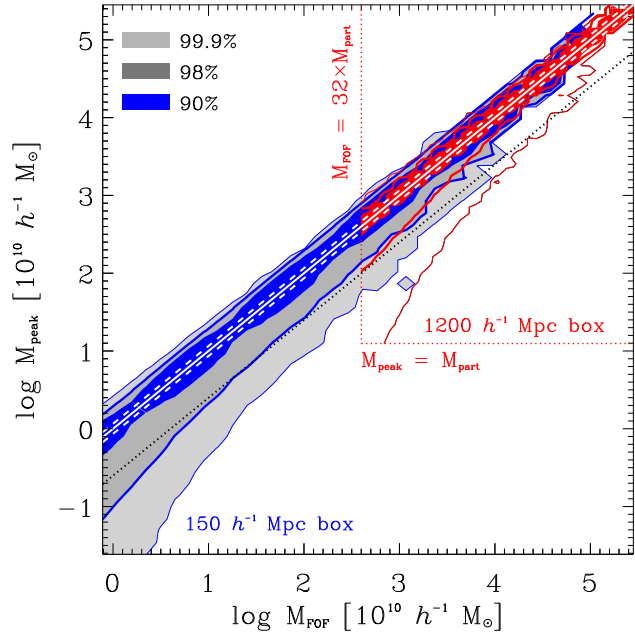}}
\end{center}
\caption{Main peak mass plotted as a function of the true friends-of-friends mass for all haloes in our cosmological simulations that contain peaks. Peak mass is determined by searching the halo particle set for peak tracers identified on the linear density field smoothed with a top-hat filter contained mass ${\rm M}_f\approx {\rm M_{FOF}}$. For cases where no such peak particle exists we scan the Lagrangian region within the radius $R_f=(3{\rm M_{FOF}}/4\pi\overline{\rho})^{1/3}$ from the proto-haloes center of mass for such a match, or identify peak tracers from smoothing scales that most accurately represent the true halo mass. For our 150\Mpch{} box the light shaded regions enclose the 99.9 percentiles of the distribution; dark shaded regions the 98$^{th}$ percentile, with filled color contours highlighting the 90$^{th}$ percentile. Results from our 1200 $h^{-1}$ Mpc box have been over-plotted using (unshaded) red contours.}
\label{Fig:MpeakMfof}
\end{figure}

\begin{figure}
\begin{center}
\resizebox{8.5cm}{!}{\includegraphics{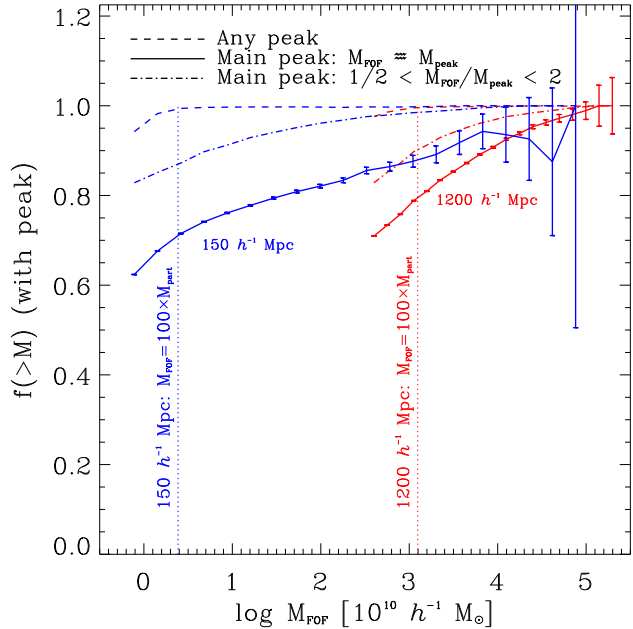}}
\end{center}
\caption{Cumulative fraction of haloes above a given mass that also contain peak particles. Dashed curves correspond to the fraction of haloes that contain peaks from {\it any} smoothing scale; solid curves with (Poisson) errors to the cumulative fraction of haloes associated with peaks of the same characteristic mass. Dot-dashed curves assume a slightly less restrictive criterion, showing the fraction of haloes whose ``main'' peak lies in the range $1/2<{\rm M_{FOF}}/{\rm M_{peak}}<2$. As in other Figures, we use blue curves to distinguish the results of our 150\Mpch{} box from those of our 1200\Mpch{} box, which are shown in red}
\label{Fig:fraction}
\end{figure}

For all haloes containing peak particles we plot the estimated main peak mass, ${\rm M_{peak}}$, versus the actual halo mass in Figure~\ref{Fig:MpeakMfof}. To avoid discreteness effects in ${\rm M_{peak}}$ we grid the data and use contours to highlight the 99.9, 98$^{th}$, and 90$^{th}$ percentiles of the distribution. For clarity, the 99.9 and 98$^{th}$ percentiles are shown as filled grey contours for our 150\Mpch, but left open for the 1200\Mpch{}-box run. Blue curves are used for the 150 $h^{-1}$ Mpc box; red curves show results for our 1200 $h^{-1}$ Mpc box. The solid white line corresponds to ${\rm M_{peak}}={\rm M_{FOF}}$, and dashed lines show the effective resolution of our sequence of smoothed density fields.

Although the scatter in ${\rm M_{peak}}$ at fixed halo mass is significant, we find that the majority of haloes can be properly associated with peaks of the same characteristic mass in the smoothed linear density field. This trend is only weakly dependent on halo mass, implying that the majority of dark matter haloes identified in cosmological simulations do, in fact, collapse around density maxima in the linear density field. This can be seen in the strong correlation between the two masses at the 90 per cent level, which follows very closely the 1:1 line anticipated by peaks theory. Finally, it is worth noting that the results of our two runs agree remarkably well at the 98$^{th}$ percentiles; even down to the $32-$particle halo limit of our group finder the two distributions are virtually indistinguishable. 

These results are shown in another way in Figure~\ref{Fig:fraction}, where we plot the cumulative fraction of haloes of mass greater than M that also contain peak particles. Different line-styles show different subsets of haloes defined in terms of the ``best'' peak mass; solid curves to the fraction of haloes for which ${\rm M_{peak}\approx M_{FOF}}$ (i.e., those for which the peak mass lies within one smoothing bin-width of the true halo mass), and dot-dashed lines to a slightly more relaxed criterion: $1/2\leq {\rm M_{FOF}/M_{peak}}\leq 2$. For comparison, we show the cumulative fraction of haloes associated with peaks from {\it any} smoothing scale with dashed lines. 

From Figure~\ref{Fig:fraction} it is clear that virtually {\it all} haloes resolved with more than 100 particles contain at least one peak particle. This is in stark contrast with the results of \citet{Katz1993a}, who found that at least 30\% of haloes in their cosmological simulations to contain no peak tracer at all. The large discrepancy between our results and those of \citet{Katz1993a} is mainly due to the fact that these authors matched haloes with peaks identified on only two smoothing scales: ${\rm M}_f\sim 2\times 10^{11} {\rm M}_{\odot}$ and $\sim 8\times 10^{11} {\rm M}_{\odot}$ (corresponding to filter volumes containing 9 and 36 particles for their simulations). Here we associated haloes with peaks on a sequence of smoothing scales spanning the entire mass range of collapsed objects in our simulations and find that $\simgt 99$\% of haloes with ${\rm N_{FOF}}>100$ particles contain at least one peak particle. In the remainder of the paper we consider only those haloes that contain at least 100 particles unless explicitly stated otherwise.

Furthermore, the majority of the haloes also contain a peak particle from the density field smoothed on the halo mass scale, implying that they form in the vicinity of peaks of the {\it same characteristic mass}. In our 150\Mpch{} box, $\sim$71\% of haloes form from such a peak. This fraction is slightly higher, at $\sim$79\% for haloes in our 1200\Mpch{} box. By scanning a range of smoothing scales for peak tracers this fraction increases; $\simgt$84\% of haloes in our 150\Mpch{} box have a peak in the smoothing range ${\rm M_{peak}}/2<{\rm M_{FOF}}<2\times {\rm M_{peak}}$, and $\sim$90\% of haloes in our 1200\Mpch{} box have such a peak. 

\begin{figure*}
\begin{center}
\resizebox{18cm}{!}{\includegraphics{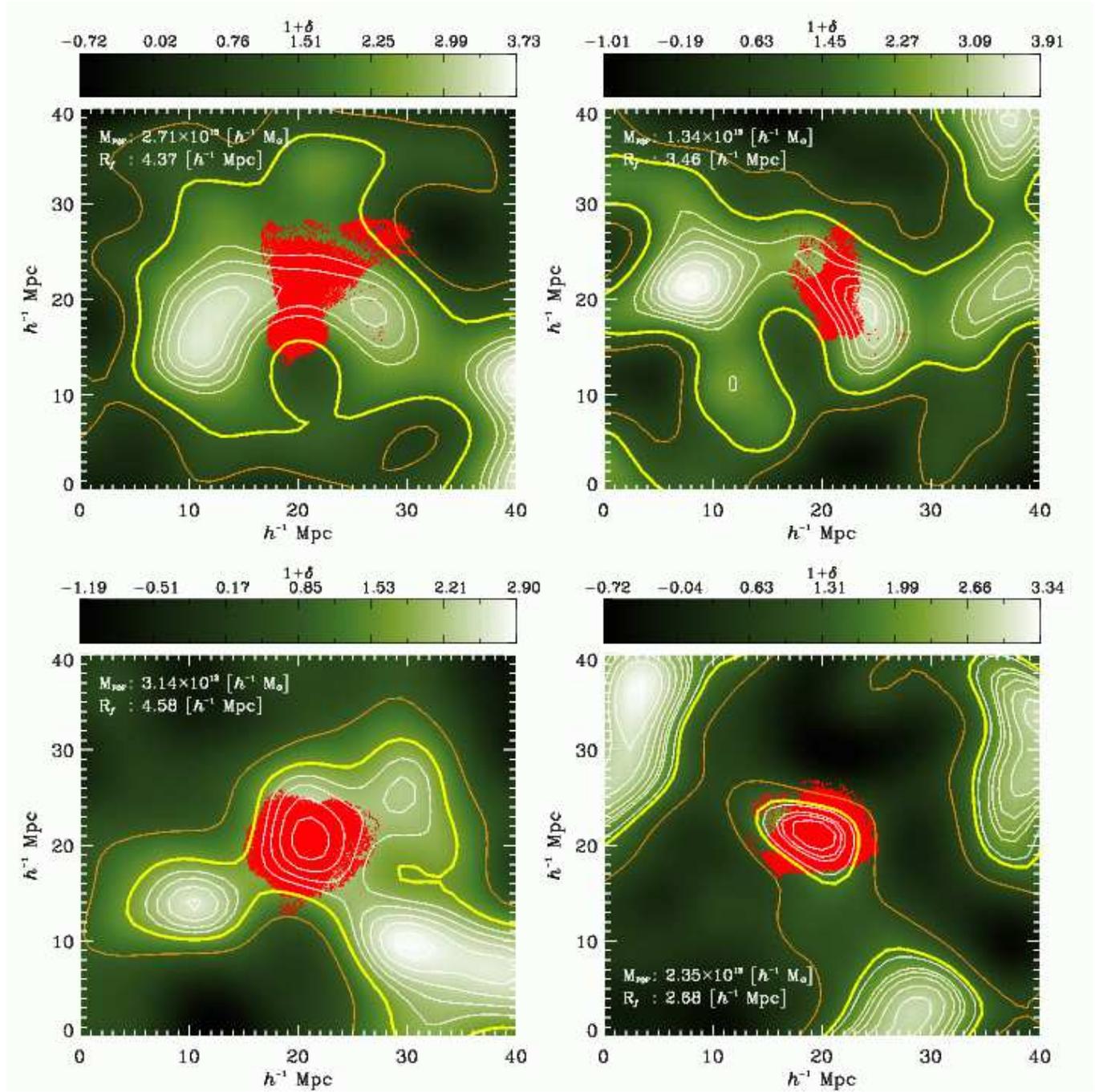}}
\end{center}
\caption{Examples of the overdensity field in the vicinity of four haloes found in our 150\Mpch{} box. Top panels show haloes for which ${\rm M_{peak}}<{\rm M_{FOF}}/4$; bottom panels show two similar mass haloes with M$_{\rm peak}\approx {\rm M_{FOF}}$. Linearly extrapolated density fields have been smoothed with a tophat filter containing mass ${\rm M_{FOF}}$. Particles that make-up the FOF haloes at $z=0$ are show as red dots. In all panels contours are used to highlight the density gradient in the neighborhood of the halo; a density contrast $\delta_s=1$ is shown as a orange curve; the threshold for spherical collapse, $\delta_s=1.686$, is shown as a thick yellow line. Note that these haloes were randomly selected from those having ${\rm M_{FOF}}\simgt 10^{13} h^{-1}$ M$_{\odot}$, and all contain more than $5\times 10^4$ particles.}
\label{Fig:density_field}
\end{figure*}

\begin{figure*}
\begin{center}
\resizebox{18cm}{!}{\includegraphics{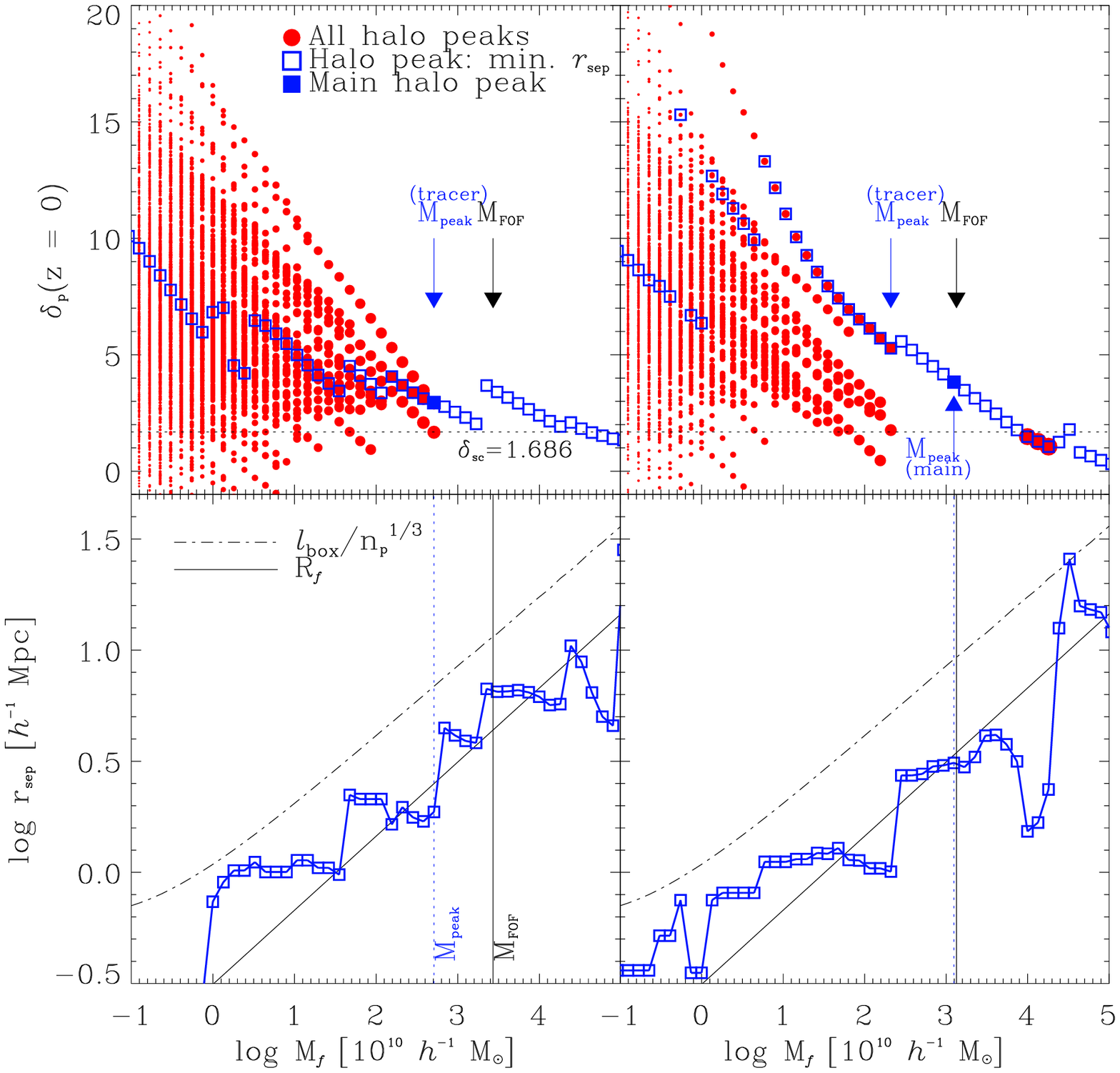}}
\end{center}
\caption{Top panels show the distribution of peak overdensities, $\delta_p$, as a function of smoothing scale for peak tracers found in the two proto-haloes shown in the upper panels of Figure~\ref{Fig:density_field}. Red points show all peak particles that, by $z=0$, are part of the FOF halo; blue squares show the peak identified on each $\delta_s$ that lies closest to the proto-halo's center of mass, regardless of its association to the halo. Downward pointing arrows mark, in each case, the halo mass, ${\rm M_{FOF}}$, and the nearest filtering mass scale for which peak tracers are found in the halo. Bottom panels show the comoving separation between the proto-halo's center of mass and the nearest peak identified on each smoothing scale. For comparison, we show the filter radius ${\rm R}_f({\rm M}_f)$ as a solid black line; the dot-dashed line shows the mean inter-peak separation as a function of smoothing scale.}
\label{Fig:peak_hist_and_sep}
\end{figure*}

\subsection{Impact of accretion geometry on estimates of ${\rm M_{peak}}$}
\label{ssec:MAH}

Although the majority of haloes in our simulations collapse near local maxima of similar characteristic mass in the linear density field, a significant fraction do not. For example, in our 150\Mpch{} box $\sim$16\% of proto-haloes contain no peak tracer in the mass range ${\rm M_{FOF}}/2<{\rm M_{peak}}<2\ {\rm M_{FOF}}$. These are {\em not} exclusively low-mass systems; of the subset of haloes with $\simgt 10^4$ particles (or a mass of $\simgt$2.43$\times 10^{12} h^{-1} {\rm M}_{\odot}$) we find that $\sim 5.6\%$ contain no peak in the quoted mass range, and a few ($\sim 1\%$) can only be matched with peaks whose mass differs from the true halo mass by at least a factor of 4. Figure~\ref{Fig:fraction} shows that similar results hold for our 1200\Mpch{} box run, in which $\sim 3.5$\% of all haloes can only be matched with linear density peaks on scales ${\rm M}_f\simlt {\rm M_{FOF}}/4$. Although these haloes do, in fact, host peak tracers we hereafter refer to them as ``peakless'' haloes for convenience (the term reflects the discrepancy between their appareent peak mass and the actual halo mass scale).

The top two panels of Figure~\ref{Fig:density_field} show examples of two peakless proto-haloes found in our 150\Mpch{} box. These were selected at random from the subset of peakless haloes that also have ${\rm M_{FOF}}>10^{13} h^{-1} {\rm M}_{\odot}$ (note that these haloes are well resolved and contain $\sim 1.1\times 10^5$ (left panel) and $\sim 5.5\times 10^4$ (right panel) particles). The choice of color scheme highlights the density contrast field, $1+\delta_s$, in the halo vicinity after smoothing on the halo mass scale. Orange curves correspond to a density contrast $\delta_s=1$, and yellow curves to the density contrast associated with the spherical collapse barrier, $\delta_{sc}=1.686$. Regions of higher density contrast are shown as thin white lines to guide the eye. For comparison, the lower two panels plot examples of two proto-haloes for which $M_{\rm peak}\approx M_{\rm FOF}$. In each panel, red dots show the projected positions of the haloes relative to the surrounding density field.

One thing to note about these projections is that the characteristic filter radii for the peakless haloes, defined as ${\rm R}_f=(3M/4\pi\overline{\rho})^{1/3}$, differ substantially from the characteristic {\it sizes} of the collapsed regions (proto-haloes have very different sizes along their principal axes). For the halo shown in the upper left-hand panel, ${\rm R}_f=4.37$\Mpch, yet the median and maximum Lagrangian distance to halo particles (measured with respect to the proto-halo's center of mass) are 4.8\Mpch{} and 14.8\Mpch{}, respectively. This suggests that the characteristic filter radius is {\it not} a good estimate of the size of proto-haloes in the linear density field. For the halo in the upper right-hand panel the situation is similar; ${\rm R}_f=3.46$\Mpch{} for this object, which is significantly smaller that the true linear dimensions of the collapsed region, which spans $\simgt 9$\Mpch. This suggests that simple spherical filtering masks the complex processes by which dark matter haloes form, and may lead to complications in associating peaks in the linear density field with collapsed objects. 

We explore these difficulties further in Figure~\ref{Fig:peak_hist_and_sep}, which examines the distribution of peak particles in and around the two peakless proto-haloes shown in the upper panels of Figure~\ref{Fig:density_field}. The top panels show, using red points, the mass dependence of peak overdensities for peak tracers belonging to each FOF halo. For comparison, open (blue) squares plot, as a function of ${\rm M}_f$, the overdensity of the peak nearest to the proto-haloes center of mass as the smoothing scale is varied, without requiring that this peak end-up in any halo by $z=0$. The distance to the nearest peak from the proto-haloes center of mass, $r_{{\rm sep}}$, is shown in the lower panel. 

Figure~\ref{Fig:peak_hist_and_sep} exposes the intrinsic difficulties with assigning peaks to collapsed objects in cosmological simulations. Consider first the results in the left-hand panels. Despite containing over $1.1\times 10^{5}$ particles, this halo hosts no peak tracers from {\it any} filtering scale ${\rm M}_f\simgt {\rm M_{FOF}}/4$. This is clearly at odds with the naive expectations of peaks theory. Further scrutiny, however, reveals the presence of a peak of the correct characteristic mass in the immediate vicinity of the proto-halo which, because of its complex accretion geometry, never becomes part of the system. Furthermore, because this peak is separated from the proto-halo by more than one filter radius it is never associated with the collapse of this object, in spite of its proximity to the proto-halo's center of mass. The ``jump'' in the value of $\delta_p$ and $r_{{\rm sep}}$ at ${\rm M_{peak}}\approx {\rm M_{FOF}}$ betrays the fact that, as ${\rm R}_f$ is increases, isolated peaks may merge with others in their vicinity which can be problematic when attempting to associate collapsed objects with peaks in $\delta_s$. 

The right-hand panels of Figures~\ref{Fig:density_field} and Figure~\ref{Fig:peak_hist_and_sep} clarify the situation. This object, with a total mass of $\sim 1.3\times 10^{13} h^{-1} {\rm M}_{\odot}$, contains no peak tracers from any smoothing scale in the mass range $2.1\times 10^{12} h^{-1} {\rm M}_{\odot} \simlt {\rm M}_f \simlt 10^{14} h^{-1} {\rm M}_{\odot}$, corresponding to a difference in peak and halo mass of at least a factor of $\sim 6$. However, by tracing the trajectory of the peak closest to the proto-halo's center, it is clear that this object does, in fact, collapse in the neighborhood of a density maxima of the appropriate characteristic mass in the smoothed density field. The upward pointing arrow marks the smoothing scale closest to ${\rm M_{FOF}}$ on which a peak lies within the characteristic filter radius of the halo's center of mass. Although this peak particle never actually ends up in the halo, it would be illusory to ignore it when attempting to map proto-haloes to peaks in the initial conditions. 

\begin{figure}
\begin{center}
\resizebox{8.5cm}{!}{\includegraphics{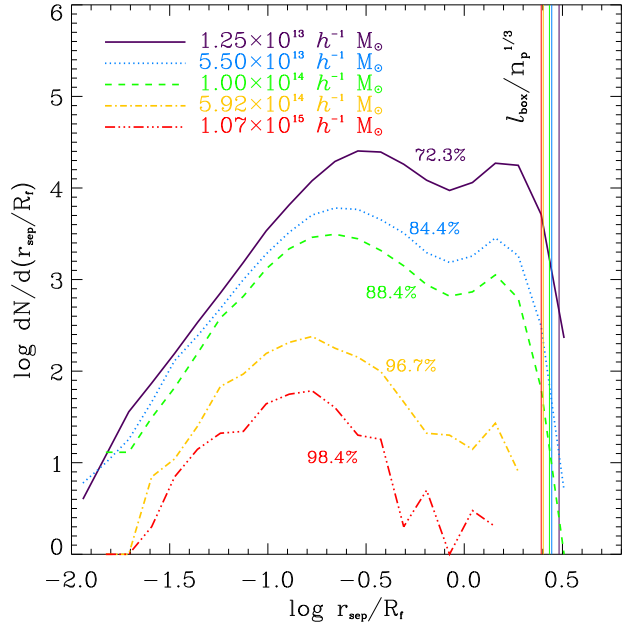}}
\end{center}
\caption{Distribution of minimum separations between proto-halo centers-of-mass and peaks of the same characteristic mass for haloes in our 1200\Mpch{} box. Different lines styles and colors correspond to haloes of different mass, which are selected to lie in a narrow mass bin of logarithmic width $\Delta\log {\rm M_{FOF}}\approx 0.129$ (equal to the logarithmic spacing between our set of smoothed linear density fields). To aid in the comparison between the various haloes samples we have normalized the separations by the characteristic filter radius for each mass. Curves are labeled by the mass-bin mid-point used to identify each set of haloes, and correspond to the filter mass-scale on which the density field was smoothed prior to identifing peaks. For each mass scale we also indicate the fraction of haloes that lie within one filter radius of the nearest peak. The vertical lines, which adopt the same color coding, show the mean inter-peak separation for each smoothed field. Clearly high mass haloes initially reside near peaks of the same mass in the linear density field, whereas for low-mass systems this is not always the case.}
\label{Fig:min_sep}
\end{figure}

This emphasizes prior suspicions that simple spherical filtering of the linear density field can be misleading, and highlights the dangers of over-interpreting the failure of previous work to provide a convincing one-to-one mapping between haloes and peaks in $\delta_s$. None the less, it is important to emphasize before proceeding that not all haloes form in the immediate vicinity of similar scale peaks. We show this in Figure~\ref{Fig:min_sep}, which plots the distribution of minimum separations between proto-halo centers and peaks {\em of the same characteristic mass}. Different colored curves correspond to haloes of different mass, selected to lie in several narrow mass bins (the bin width, in each case, coincides with the smoothing bin width for our sequence of smoothed density fields). We show only results for haloes in our 1200\Mpch{} box. For each mass bin, the percentage of haloes whose nearest peak lies within one filter radius is indicated. As expected from Figure~\ref{Fig:fraction}, high mass haloes virtually always reside near peaks of the same characteristic mass. Towards lower masses, however, the distribution of minimum separations becomes increasingly bimodal; one ``hump'' corresponds to the majority of haloes that reside near peaks initially, the other is induced by the boundary imposed by the mean inter-peak separation. The vertical lines in Figure~\ref{Fig:min_sep} show the average separation between peaks on each mass scale and clearly show that the vast majority of haloes lie within this characteristic distance scale from their nearest peak, even if they are physically unassociated with it.

In the next section, we examine some of the statistical properties of peakless haloes, which we contrast directly with the bulk of the halo population. To this end, we divide the sample of haloes into two subsets: those clearly associated with peaks in $\delta_s$, for which ${\rm M_{peak}\approx M_{FOF}}$, with which we directly contrast those whose best peak mass is ${\rm M_{peak}}\simlt {\rm M_{FOF}}/4$. 

\begin{figure}
\begin{center}
\resizebox{8.5cm}{!}{\includegraphics{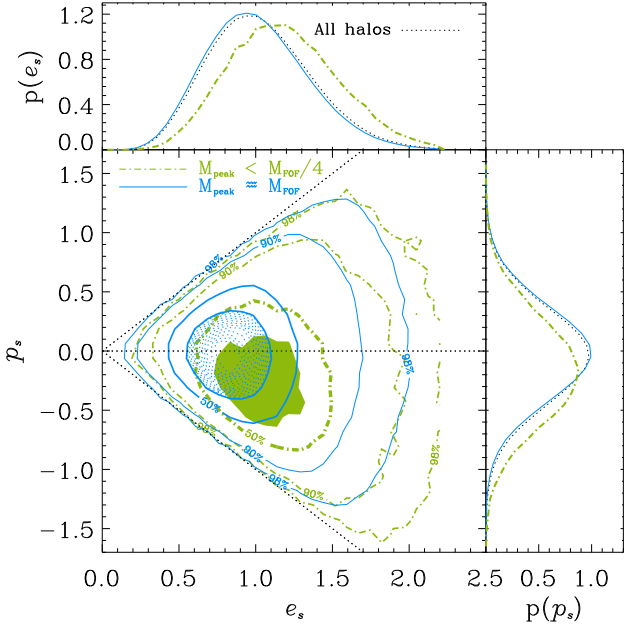}}
\end{center}
\caption{Prolateness and ellipticity of the velocity shear tensor measured at proto-halo centers in both of our simulations. Only haloes with ${\rm N}>100$ particles are shown. Blue contours show the distribution for proto-haloes properly matched with peaks in the density field when smoothed on the same characteristic scale. Green contours show the corresponding distributions, but for proto-haloes having M$_{\rm peak}\simlt {\rm M_{FOF}}/4$. The corresponding probability distributions are shown in the upper and right-hand panels.}
\label{Fig:ellip_prol}
\end{figure}

\begin{figure}
\begin{center}
\resizebox{8.5cm}{!}{\includegraphics{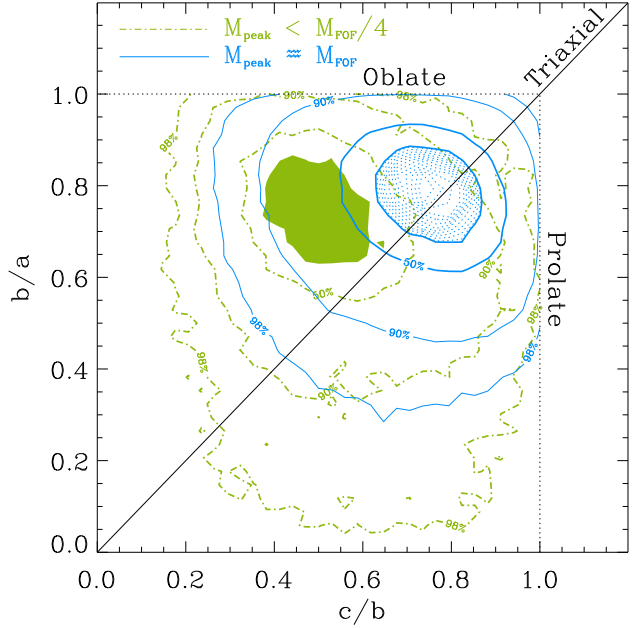}}
\end{center}
\caption{Intermediate-to-major ($b/a$) versus minor-to-intermediate ($c/b$) axis ratios for the same subsamples of proto-haloes plotted in Figure~\ref{Fig:ellip_prol}. Perfectly oblate haloes have $b/a=1$, while perfectly prolate objects have $c/b=1$; ``maximally triaxial'' haloes have $b/a=c/b$. Blue contours highlight the regions occupied by the subsample of haloes having ${\rm M_{peak}\approx M_{FOF}}$; green contours show the proto-haloes with ${\rm M_{peak}}\simlt {\rm M_{FOF}}/4$.}
\label{Fig:shapes}
\end{figure}

\subsection{The importance of velocity shear}
\label{ssec:shear}

Another point worth mentioning is that both haloes shown in the upper panels of Figure~\ref{Fig:density_field} appear to collapse along high-density ridges that connect massive structures in the linear density field. This is, in fact, a common occurence for peakless haloes found in our simulations, and is not unique to the examples plotted in Figure~\ref{Fig:density_field}. The proto-halo shown in the upper-left panel of Figure~\ref{Fig:density_field} is initially elongated in the direction transverse to the density ridge. This is consistent with the ellipsoidal collapse model in which maximum compression occurs along the primary axis of the shear field, which is presumably orthogonal to the ridge.

A possible implication is that the external tides associated with the surrounding large scale structure may have a non-negligible impact on the collapse and subsequent virialization of these systems \cite[e.g.][]{Lilje1986}. As discussed by \citet{vandeWeygaert1994} \cite[see also][]{Hoffman1986,Bertschinger1994}, a strong external shear is able to accelerate the growth of a small scale fluctuations leading to the formation of massive haloes in the absence of a corresponding peak of similar scale in the linear density field. This is because convergent velocity flows induced by large scale structure may enhance the collapse of nearby small scale objects, an effect which may play an important role in the formation of low mass haloes \citep{Hoffman1986,Hoffman1989}. Tidal forces are also capable of altering the angular momentum acquisition of proto-haloes \cite[e.g.][]{White1984,Porciani2002a}, as well as their kinematic and spatial distribution \cite[e.g.][]{Binney1979,Dubinski1992}. 

We quantify the linear tidal field at proto-halo centers in terms of the deformation tensor
\begin{equation}
D_{ij}=\frac{\partial^2\Phi}{\partial x_i\partial x_j}.
\label{eq:deformation}
\end{equation}
Since the trace of $D_{ij}$ is equal to $\delta$, we define the {\em velocity shear} as its traceless part, $T_{ij}=D_{ij}-(D_{ii}/3)\delta_{ij}$. Here $\delta_{ij}$ is the Kronecker delta (not to be confused with the density contrast), and $\Phi({\mathbf{x}})$ is the peculiar gravitational potential which is related to $\delta$ through Poisson's equation,
\begin{equation}
\nabla^2\Phi=\delta.
\label{eq:potential}
\end{equation}

In general, the collapse time depends not only on the trace of $D_{ij}$ (i.e., on $\delta$) but rather on its three eigenvalues, which determine the growing mode perturbations: local maxima in $\delta$ are {\it not} necessarily local minima in collapse time. We quantify the tidal field at a given point in terms of the eigenvalues of $D_{ij}$, $\lambda_1\geq\lambda_2\geq\lambda_3$, which determine whether the initial flow around a given point is of compression or dilation. Equivalent quantities are the shear ellipticity, $e$, prolaticity, $p$, and density contrast, $\delta$. For $\delta > 0$, we have $e\geq 0$ and $-e\leq p\leq e$; the relationship to the eigenvalues of the tidal tensor are conventionally given by (e.g. BBKS)
\begin{equation}
e = \frac{\lambda_1-\lambda_3}{2\delta},
\label{eq:ell}
\end{equation}
and
\begin{equation}
p = \frac{\lambda_1+\lambda_3-2\lambda_2}{2\delta}.
\label{eq:prl}
\end{equation}

It is customary to quantify the {\em velocity} shear at a given point in terms of the modified eigenvalues: $t_i=\lambda_i-\delta/3$. The sign of the $t_i$'s determine whether local flows promote collapse around a given point (positive), or impede it (negative). Since $\sum_i t_i =0$, we may also characterize the velocity shear in terms of its ellipticity, $e_s$, and prolateness, $p_s$, where
\begin{equation}
e_s = \frac{t_1-t_3}{\sigma_0} \textrm{,  and  }  p_s = \frac{3(t_1+t_3)}{\sigma_0},
\label{eq:ellprls}
\end{equation}
and $\sigma_0$ is given by eq.~\ref{eq:spectralmoments}. The magnitude of $p_s$ determines the relative contributions from velocity shear along the first and third axis of the flow; the sign of $p_s$ indicates whether the flow along the intermediate axis is inward ($p_s<0$) or outward ($p_s>0$).

In Figure~\ref{Fig:ellip_prol} we plot the distribution of the ellipticity, $e_s$, and prolaticity, $p_s$, of the velocity shear measured at the centers of mass of two subsamples of proto-haloes after stacking all (${\rm N_{FOF}}>100$) haloes in both of our simulations. For all haloes $e_s$ and $p_s$ are measured from the tidal field smoothed on the mass scale of the halo (note that we have normalized by the mass variance, $\sigma_0({\rm M})$, to minimize any subtle mass dependence in the trends). Blue curves show the results for all proto-haloes properly identified with peaks of the same mass in the linear density field; green contours to peakless haloes. Curves mark the isodensity contours which enclose the 98$^{th}$, 90$^{th}$, 50$^{th}$, and 25$^{th}$ percentiles of the distribution. The top and side panels show the corresponding probability distributions for $e_s$ and $p_s$. 

The velocity shear measured at proto-halo centers is slightly but significantly different for the two halo subsamples. Peakless haloes have slightly more negative prolaticities implying that, on average, $t_2\simgt 0$ (since $t_2$ carries the opposite sign to $p_s$). Recalling that $\sum_i t_i=0$, and $t_3\leq 0$, this implies that the magnitude of the velocity outflow along the direction of $t_3$ is, on average, larger than the magnitude of the velocity inflow along the primary and intermediate axes of the shear field, i.e. $|t_3|\simgt |t_1|$. This indicates a slightly higher tendency for the velocity shear to aid in the collapse of these density peaks along two, rather than one axis of the shear field. Considering all proto-haloes with ${\rm N_{FOF}}>100$ particles, we find that $\sim$52\% of those with ${\rm M_{peak}}\approx {\rm M_{FOF}}$ have $t_2>0$, whereas $t_2>0$ for $\sim$64\% of peakless haloes.

The differences in velocity shear for the two subsamples of haloes highlighted in Figure~\ref{Fig:ellip_prol} suggests that the geometry of the Lagrangian region from which these haloes grow may also differ systematically. For example, a halo with $t_1 > t_2 > 0$ is expected to form from a flatter initial configuration than a similar mass object with $t_1>0$ and $t_2<0$, since in the latter case the velocity shear impedes collapse along the intermediate axis rather than aids it.

We show this explicitly in Figure~\ref{Fig:shapes}, which plots the intermediate-to-major ($b/a$) versus minor-to-intermediate ($c/b$) axis ratios for the same subsamples of proto-haloes used above. (Axis ratios are computed from the eigenvalues of the proto-halo's intertia tensor after weighting each particle position by its distance from the center of mass.) Note that a perfectly prolate spheroid has $c/b=1$, whereas an oblate spheroid corresponds to $b/a=1$; a maximally triaxial object has $b/a=c/b$. Contours highlight the same iso-density regions as those in Figure~\ref{Fig:ellip_prol}. It is clear that proto-haloes with ${\rm M_{peak}}\approx {\rm M_{FOF}}$ (which accounts for more than three-quarters of all haloes) evolve from initial Lagrangian regions whose geometries are on average triaxial, with no obvious preference for prolate or oblate configurations. However, as anticipated from Figure~\ref{Fig:ellip_prol}, peakless proto-haloes clearly deviate from the average, showing a much stronger tendency to evolve from initially more flattened configurations, suggesting that these halo subsets experience very different accretion geometries. Since the eigenvalues of the tidal field and proto-halo intertia tensor are nearly perfectly aligned \citep{Lee2000,Porciani2002b,Lee2009}, we speculate that the initial proto-halo shape is dictated by the large scale tidal field. We will address this issue in forthcoming work (Ludlow \& Porciani, 2010, in prep.).

\subsection{Assembly bias}
\label{ssec:assemblybias}

\begin{figure}
\begin{center}
\resizebox{8.5cm}{!}{\includegraphics{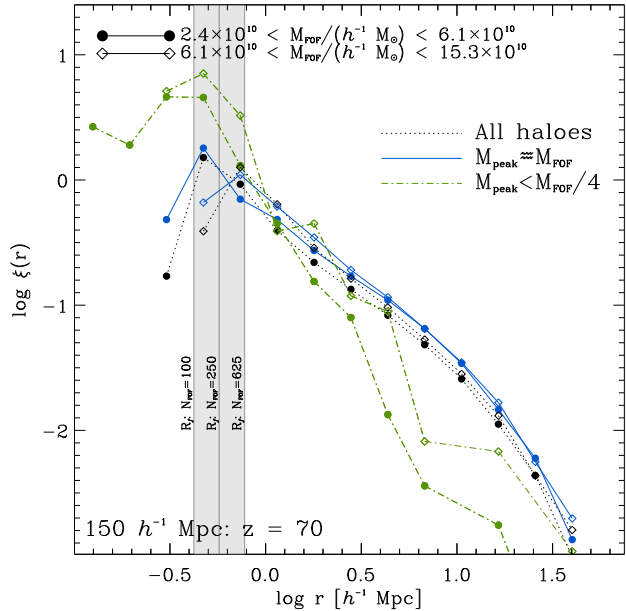}}
\end{center}
\caption{Two-point correlation function for proto-haloes spanning two separate mass ranges: $2.4\times 10^{10} h^{-1}{\rm M}_{\odot}<{\rm M_{FOF}}<6.1\times 10^{10} h^{-1}{\rm M}_{\odot}$ (solid circles), and $6.1\times 10^{10} h^{-1}{\rm M}_{\odot}<{\rm M_{FOF}}<15.3\times 10^{10} h^{-1}{\rm M}_{\odot}$ (open diamonds). These mass ranges include haloes having between 100 and 250 particles, and between 250 and 625 particles, respectively. All correlation functions are computed at the initial redshift of our 150\Mpch{}-box simulation using proto-halo centers-of-mass. The dotted black curves show $\xi(r)$ for all haloes in each mass bin. The solid blue curves distinguish haloes for which ${\rm M_{peak}} \approx {\rm M_{FOF}}$ from peakless haloes, which are shown using dot-dashed green lines. For comparison, the spread in filter radii for haloes in each mass range is shown as shaded vertical bands.}
\label{Fig:xi_prof_zIC}
\end{figure}

\begin{figure}
\begin{center}
\resizebox{8.5cm}{!}{\includegraphics{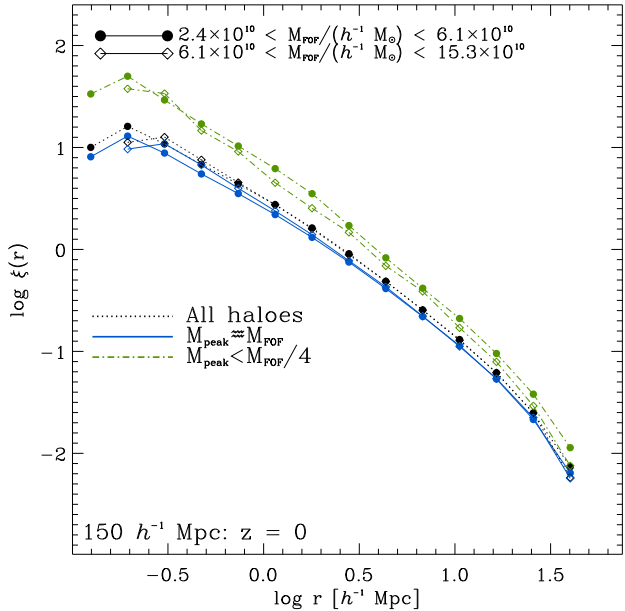}}
\end{center}
\caption{Auto-correlation functions for the same halo subsample used in Figure~\ref{Fig:xi_prof_zIC}, but plotted at $z=0$. Line-styles and colors have the same meaning as before. Note the excess clustering of peakless haloes on scales $\simgt 1-2\Mpch$.}
\label{Fig:xi_prof_z0}
\end{figure}

\cite{ShethTormen2004}, \cite{Gao2005} and subsequent studies have conclusively demonstrated that the clustering strength of dark matter haloes depends on parameters other than their mass. Halo properties such as formation time, concentration, spin, shape, internal dynamics, and substructure content all show correlations with the strength of clustering \cite[e.g.][]{Wechsler2006,Croton2007,Bett2007,Gao2007,Faltenbacher2010}. Somewhat suprisingly, it has been noted that the dependence of clustering on these parameters does not follow intuitively from the intrinsic parameter correlations. For example, older haloes are on average more concentrated than younger ones, but the dependence of clustering on formation time and concentration exhibit an opposite behaviour. This implies that the clustering properties of dark matter haloes may depend on a more fundamental parameter, or on a number of aspects of the variety of halo formation histories.

In Section~\ref{ssec:masscomp} we identified two classes of haloes: those clearly associated with similar scale peaks in the linear density field, and those with no such peak. We have shown that the velocity shear and proto-halo shape for these two subsamples are systematically different; peakless haloes are more strongly sheared and evolve from more oblate initial configurations than the average halo. The possibility that these differences may be influenced by environment motivates us to explore the statistical properties of their spatial distributions by means of the two-point correlation function, $\xi (r)$. 

Figure~\ref{Fig:xi_prof_zIC} shows $\xi(r)$ for proto-halo centers measured at the initial redshift of one our simulations. To avoid mixing results from haloes of widely different masses we plot here only haloes found in our 150\Mpch{} box, that lie in two narrow mass bins: $2.4\times 10^{10}<{\rm M_{FOF}}/(h^{-1}{\rm M}_{\odot})<6.1\times 10^{10}$ (shown as connected dots in Figure~\ref{Fig:xi_prof_zIC}), and $6.1\times 10^{10}<{\rm M_{FOF}}/(h^{-1}{\rm M}_{\odot})<15.3\times 10^{10}$ (connected diamonds). These mass bins correspond to haloes having between 100 and 250 particles, and between 250 and 625 particles, respectively. The dotted line shows the auto-correlation function of all proto-haloes in each mass bin; blue and green curves show the effect of limiting the calculation of $\xi(r)$ to haloes for which ${\rm M_{peak}}\approx {\rm M_{FOF}}$, and those with ${\rm M_{peak}}< {\rm M_{FOF}}/4$, respectively. The separations plotted on the horizontal axis are in co-moving units.

Clearly these halo subsamples exhibit very different initial clustering properties. On small scales ($\simlt 1$\Mpch) peakless haloes are significantly more clustered than average, but the trend is reversed for larger separations. The correlation function of peakless haloes also has a different slope than that of the full halo sample. Between $\sim 2-10\Mpch{}$, for example, the amplitude of the correlation functions can differ by over an order of magnitude. Although this effect is strongest in the lowest halo mass bin it is still clearly present in both, suggesting that it is not particularly sensitive to the selected halo mass range. We have noted, however, that a small but significant fraction of peakless haloes tend to reside in the immediate vicinity of more massive systems in our initial conditions. For example, $\sim$10\% of peakless haloes in the mass range $2.4\times 10^{10}<{\rm M_{FOF}}/(h^{-1}{\rm M}_{\odot})<6.1\times 10^{10}$ lie within the characteristic radius R$=(3{\rm M}/4\pi \overline{\rho})^{1/3}$ of a halo with M$\simgt 10^{13.5}\Mpch{}$; for those halos with ${\rm M_{peak}}\approx {\rm M_{FOF}}$ this fraction is around $3\%$. Excluding these halos from the calculation of $\xi(r)$ results in a significant change in clustering strength of peakless halos on scales $\simgt 3\Mpch{}$, which then more closely follow the blue curves in Figure~\ref{Fig:xi_prof_zIC}. 

In Figure~\ref{Fig:xi_prof_z0} we plot the $z=0$ two-point auto-correlation functions for the same halo subsamples, using the same linestyles and colors as in Figure~\ref{Fig:xi_prof_zIC}. At the smallest separations ($r\approx 2$\Mpch), the difference in clustering strength for the two subsamples differs by a factor of $\sim \ 4$. Combined with the results on the shapes of proto-haloes in Figure~\ref{Fig:shapes}, this suggests an intimate connection between the accretion geometry of a given halo and its proximity to neighboring systems \cite[see also,][]{WangMoJing2007,Hahn2009}. Surprisingly, we find that the $z=0$ correlation function of peakless haloes now exceeds that of haloes with ${\rm M_{peak}}\approx {\rm M_{FOF}}$ at {\em all} separations. This is particularly intriguing since, on linear scales, the Eurlerian bias $b_E$ should be related to the Lagrangian bias, $b_L$, by $b_E=b_L+1$ \citep{MoWhite1996}. Further study aimed at addressing the physical origin of this behaviour is currently ongoing.

Although the results presented in Figures~\ref{Fig:xi_prof_zIC} and \ref{Fig:xi_prof_z0} are limited to haloes in a narrow mass range, and only one cosmological simulation, they appear robust to the specifics of this selection criterion. The fact that the initial shape of proto-haloes, and the strength by which they are sheared, is related to their clustering strength may have interesting implications for galaxy formation models.

\section{Summary}
\label{sec:conclusions}

We have used two fully cosmological simulations of structure growth in the standard $\Lambda$CDM cosmogony to study the conditions from which dark matter haloes emerge from the linear density field. The primary focus of this work was to perform stringent yet simple tests of the central ansatz of the peaks formalism, which states that dark matter haloes of mass M evolve from peaks in the linear density field when the latter is smoothed with a filter of the same characteristic mass. 

We identify possible sites for halo collapse by identifying local maxima, or peaks, on a sequence of smoothed density fields that span the entire mass range of FOF haloes in our simulations. For each smoothed density field we build a list of peak particles by tagging those nearest to each peak grid point, which we then cross-correlate with the halo particles in our friends-of-friends catalogs. Our main results can be summarized as follows.

\begin{itemize}

\item We define the ``main'' halo peak as the one identified on the smoothed field for which M$_f$ is closest to the true halo mass, and use this to ``predict'' the expected mass of the halo. We find that as many as $\sim$70\% of {\em all} haloes (with ${\rm N_{FOF}}>100$) in both of our simulations can be properly identified with peaks in the linear density field when smoothed on the mass scale of the halo. This fraction depends systematically on halo mass, with as many as $\sim$91\% of haloes with M$> 5\times 10^{14} h^{-1} {\rm M}_{\odot}$ forming in the vicinity of peaks of the expected characteristic mass. By scanning each halo for peaks over a broader range of smoothing scales we find better agreement: more than $\sim 85\%$ of haloes form from peaks in the mass range ${\rm M}_{\rm peak}/2 \simlt {\rm M}_{\rm FOF}\simlt 2\times {\rm M}_{\rm peak}$.

\item Although the majority of haloes in our simulations form preferentially around similar-scale peaks in the linear density field a small but significant fraction show a considerable disparity between the predicted and measured masses. For example, $\sim$20\% of haloes with $>100$ particles have ${\rm M}_{\rm peak} < {\rm M}_{\rm FOF}$, of which $\sim 15$\% contain no peak tracers from any scale within a factor of four of the true halo mass. These are typically low-mass objects, but several have ${\rm M}\simgt 10^{13} h^{-1} {\rm M}_{\odot}$ and contain more than $10^4$ particles. Haloes for which ${\rm M}_{\rm peak}\simlt {\rm M}_{\rm FOF}/4$ we have referred to as ``peakless'' haloes for convenience.

\item By contrasting directly the properties of peakless proto-haloes with those of the population bulk we find slight but systematic differences in both the linear tidal shear and initial shapes of the two samples. The tidal shear (measured at proto-halo centers, and smoothed on the halo mass scale) acting on the proto-haloes implies that, on average, peakless haloes are tidally compressed along two axes of the initial tidal tensor, and expanding along one. 
On the other hand, 
 those with ${\rm M_{peak}}\approx {\rm M_{FOF}}$ are characterised by roughly equal compression and expansion along two axes of the tidal field, and a null flow along the third.

\item Peakless haloes also evolve from more oblate initial configurations than the nearly triaxial geometries characterising the mean. Haloes for which ${\rm M_{peak}}\approx{\rm M_{FOF}}$ typically collapse from Lagrangian regions which are initially triaxial with typical axis ratios $b/a\approx c/b\approx 0.75$. Peakless haloes, on the other hand, are initially more flattened, having $b/a\approx 0.75$ and $c/b\approx 0.5$. The differences in the initial shapes of proto-haloes are qualitatively consistent with the differences in their tidal shear; we plan to address the interrelationship between tidal forces and accretion geometry in a forthcoming paper (Ludlow \& Porciani, 2010, in prep.).

\item The auto-correlation functions of normal proto-haloes and of peakless haloes show significant differences at both the initial redshift of the simulation as well as at $z=0$; on scales $\simlt 1\Mpch{}$ peakless haloes are significantly more clustered both initially as well as today. The evolution of the correlation functions for the different subsamples, however, appears quite different; on scales larger than a few Megaparsecs peakless haloes are today {\em more} clustered than average, whereas they are initially {\em less} clustered at the same separations. Further study is required to address the origin of this behaviour. Combined with the differences in the initial shapes and shear properties of these samples, these results imply that the accretion geometry (and hence the mass acquisition history) of highly clustered haloes can differ substantially from that of field haloes. This may have important implications for understanding the physics of galaxy formation in dense environments and for understanding the role environment plays in the morphological transformation of galaxies. 

The main goal of this work has been to provide a simple and comprehensive test of the central ansatz of the peaks formalism. In doing so we have demonstrated that the majority of dark matter haloes identified in our cosmological simulations form in the vicinity of linear density peaks of the expected characteristic mass in the simulations initial conditions, yet a significant fraction do not. Our work sparks several interesting questions for future study: what is the physical origin of the peakless haloes? What implications do these objects have for analytic models of structure formation rooted in peaks theory? What is the nature of their clustering properties and their imprint on the assembly bias? We plan to address these and other pertinent issues in forthcoming work.

\end{itemize}

\section*{Acknowledgements}
We would like to thank the referee, Vincent Desjacques, for a constructive
report that improved the clarity of the paper. We also acknowledge partial 
financial support through the SFB-Transregio 33 ``The Dark Universe'' by 
the Deutsche Forschungsgemeinschaft (DFG).

\bibliographystyle{mn2e}
\bibliography{paper}

\end{document}